\documentclass[prb, aps, twocolumn, superscriptaddress,noshowpacs]{revtex4}
\usepackage{eqnarray}
\usepackage[english]{babel} 
\usepackage[english]{layout}
\usepackage{amsmath,amsfonts,amssymb}
\usepackage{graphicx}
\usepackage{float}
\usepackage{color}
\usepackage{braket}
\usepackage{version}
\usepackage{array,multirow,makecell}

\begin{document}
\title{Competition between horizontal and vertical polariton lasing in planar microcavities}
\author{O. Jamadi}
\affiliation{Universit\'e Clermont Auvergne, CNRS, SIGMA Clermont, Institut Pascal, F-63000 Clermont-Ferrand, France.} 
\affiliation{Univ. Lille, CNRS, UMR 8523 -- PhLAM -- Physique des Lasers Atomes et Molécules, F-59000 Lille, France} 
\author{F. R\'everet}
\author{D. Solnyshkov}
\author{P. Disseix}
\author{J. Leymarie}
\affiliation{Universit\'e Clermont Auvergne, CNRS, SIGMA Clermont, Institut Pascal, F-63000 Clermont-Ferrand, France.} 
\author{L. Mallet-Dida}
\author{C. Brimont}
\author{T.~Guillet}
\affiliation{Laboratoire Charles Coulomb (L2C), Universit\'e de Montpellier, CNRS, Montpellier, France}
\author{X. Lafosse}
\author{S. Bouchoule}
\affiliation{Centre Nanosciences et Nanotechnologies (C2N), CNRS/UPSUD, Avenue de la Vauve, Palaiseau, France.}
\author{F. Semond}
\author{M. Leroux}
\author{J. Zuniga-Perez}
\affiliation{UCA, CRHEA-CNRS, Valbonne, France} 
\author{G. Malpuech}
\affiliation{Universit\'e Clermont Auvergne, CNRS, SIGMA Clermont, Institut Pascal, F-63000 Clermont-Ferrand, France.}

\begin{abstract}
Planar microcavities filled with active materials containing excitonic resonances host radiative exciton-polariton (polariton) modes with in-plane wave vectors within the light cone. They also host at least one mode guided in the cavity plane by total internal reflection and which is not radiatively coupled to the vacuum modes except through defects or sample edges. We show  that polariton lasing mediated by polariton stimulated scattering can occur concomitantly in both types of modes in a microcavity. By adjusting the detuning between the exciton and the radiative photon mode one can favor polariton lasing either in the radiative or in the guided modes. Our results suggest that the competition between these two types of polariton lasing modes may have played a role in many previous observations of polariton lasing and polariton Bose Einstein condensation.
\end{abstract}
\maketitle

\section{Introduction}

Exciton-polaritons (polaritons) are quasi-particles resulting from the coupling between a light mode and an excitonic resonance. They have been theoretically introduced by Hopfield and Agranovich \cite{Hopfield1958, Agranovich1959} at the end of the 50's to describe light propagation in bulk semiconductors. In 1992, the achievement of strong light-matter coupling between the photonic radiative modes (within the light cone) of a planar microcavity and the excitonic resonances of embedded quantum wells \cite{Weisbuch1992} opened the era of 2-dimensional cavity polaritons, which since then have been very extensively studied both from a fundamental and from an applied point of view \cite{Microcavities,Sanvitto2016}. In a microcavity the radiative mode is not, however, the only photonic mode expected to cross the exciton energy. Indeed, at least one mode, fully guided in the plane of the cavity by total internal reflection, is expected to give rise to a guided polariton. The existence of these modes was perfectly understood in the 90's, their presence being responsible for the relatively weak Purcell enhancement factors \cite{Yamamoto1991} of planar cavities. In the first attempts to reach the polariton lasing regime following the proposal of Imamoglu and Ram \cite{Imamoglu1996}, they were correctly identified as "leaky modes". 

The concept of polariton laser is an extension of the concept of exciton laser formulated in the 60s \cite{Haug1967}. Exciton laser in a bulk semi-conductor is based on the scattering of an exciton-like polariton either on another exciton-like polariton or on a LO-phonon towards a polariton state with a lower energy and, thus, with a larger photonic fraction. If this relaxation process is more efficient than the reverse one, one can speak of an excitonic gain. In bulk material, polariton states are stationary propagating states which are not radiatively coupled to the continuum of vacuum photonic modes existing outside of the sample.  Light coming from these modes leaks out only from the edge of the sample or through sample defects which provoke Rayleigh scattering.  The polariton laser proposal \cite{Imamoglu1996} added a vertical cavity on top of the exciton laser, where the modes to be amplified were the radiative vertical polariton modes of the cavity. Contrary to bulk polariton states, they are not intrinsically propagating modes, being at rest at zero in-plane wave vector. They moreover possess a finite radiative lifetime. Polariton lasing requires both the excitonic gain to be present and the spontaneous scattering rate to be faster than the polariton decay, this last condition being absent in the simpler exciton laser picture. Polariton lasing and the related polariton Bose-Einstein Condensation in planar cavities have then been reported in a large amount of cavity systems \cite{Kasprzak2006,Christopoulos2007,Christmann2008,Balili07,Bajoni08,Cohen2010,Feng2013,Plumhof2014,Dietrich2016,Sun2017,Su2017},  becoming a very intense research topic. On the other hand, the possibility of polariton lasing in guided polariton modes \cite{Liscidini2011,Pirotta2014,Lerario2017,Walker2013,Walker2017,Rosenberg2016,Ellenboger2011,Ciers2017,Lagoudakis2017,Hu2017,Jamadi2017}
  has been considered only recently \cite{Solnyshkov2014}, to our knowledge. In this last case, one needs to define certain system boundaries, which allow to introduce the notion of mean time spent by the polaritons in the gain region. The first observation of  polariton lasing based on guided modes \cite{Jamadi2017}, performed in a monolithic ZnO-based waveguide, has shown pumping thresholds comparable to the ones obtained for vertical polariton lasing in state-of-the-art ZnO-based cavities. This similarity can be understood by simple estimates of the traveling time of guided polaritons under a typical pumping spot 10-20 $\mu$m wide, which is of the order of 1 ps. This is comparable with the polariton radiative lifetime in many types of vertical cavities where polariton lasing has been demonstrated. It suggests that polariton leakage into guided modes and even horizontal polariton lasing could be seriously competing with vertical polariton lasing in a large part of experimental configurations. 

In this work, we study a ZnO-based microcavity showing a Q-factor of 2000 \cite{Jamadi2016}.
Under non-resonant optical pumping we are able to observe simultaneously polariton lasing under the pumping spot, associated with the vertical radiative modes, and also, at a different energy, polariton lasing in the guided polariton modes, essentially extracted through the sample (mesa) edges. By changing either the pumping spot size or the detuning between the exciton and the radiative photon mode, we can tune the vertical cavity threshold and favor one type of polariton lasing with respect to the other. 

\section{Sample description and experimental setup}

The sample under study \cite{Zuniga2014} consists of a ZnO active layer embedded between a 30-pair AlN/AlGaN bottom distributed Bragg reflector (DBR) and an 11-pair SiO$_2$/HfO$_2$ top DBR. The cavity thickness is 290 nm (2 $\lambda$) and the Rabi splitting 160~meV. The microcavity structure is grown on a silicon mesa having a lateral size of 100~$\mu$m. The presence of lateral free surfaces (through which guided modes can be extracted) is essential for our observations. As studied in detail in Ref.~\cite{Jamadi2016}, this thick cavity supports several radiative modes being relatively close in energy. It also supports several fully guided modes which should strongly couple to exciton resonances and that have not been considered previously. The sample is studied through micro-photoluminescence (PL) using the fourth harmonic (266 nm) of a Nd:YAG laser as the excitation source. The pulse duration is 400 ps and the repetition rate is 20 kHz. A UV microscope objective with 0.5 numerical aperture was used to obtain small excitation spots of different diameters. The emission of the sample was collected through the same objective and imaged onto the spectrometer slit by a spherical lens mounted on a motorized translation stage, which allows to collect emission from various areas of the sample. All the measurements are performed at 5~K.

Figure 1(a) shows a realistic sketch of the polariton dispersion computed using a model of three coupled oscillators (dotted lines) representing the exciton resonance, a cavity mode dispersion, and the fundamental guided TE mode of energy $E=\hbar c k/n$, where $k$ is the in-plane wave vector, $c$ the speed of light, and $n$ the effective refractive index of the cavity material. For simplicity, we consider only one cavity mode and one guided mode. The solid black line shows the light cone where the radiative emission in vacuum is possible. The radiative and guided polariton modes are populated by the same relaxation processes represented by red arrows,  which are assisted by exciton-phonon, exciton-exciton, and exciton-free carrier scattering mechanisms \cite{Microcavities}. Polaritons can scatter from the guided to the radiative polariton branch and vice versa by elastic scattering on static disorder, which typically occurs on fast sub-picosecond time scales.

\begin{figure}[tbp]
 \begin{center}
 \includegraphics[width=1.0\linewidth]{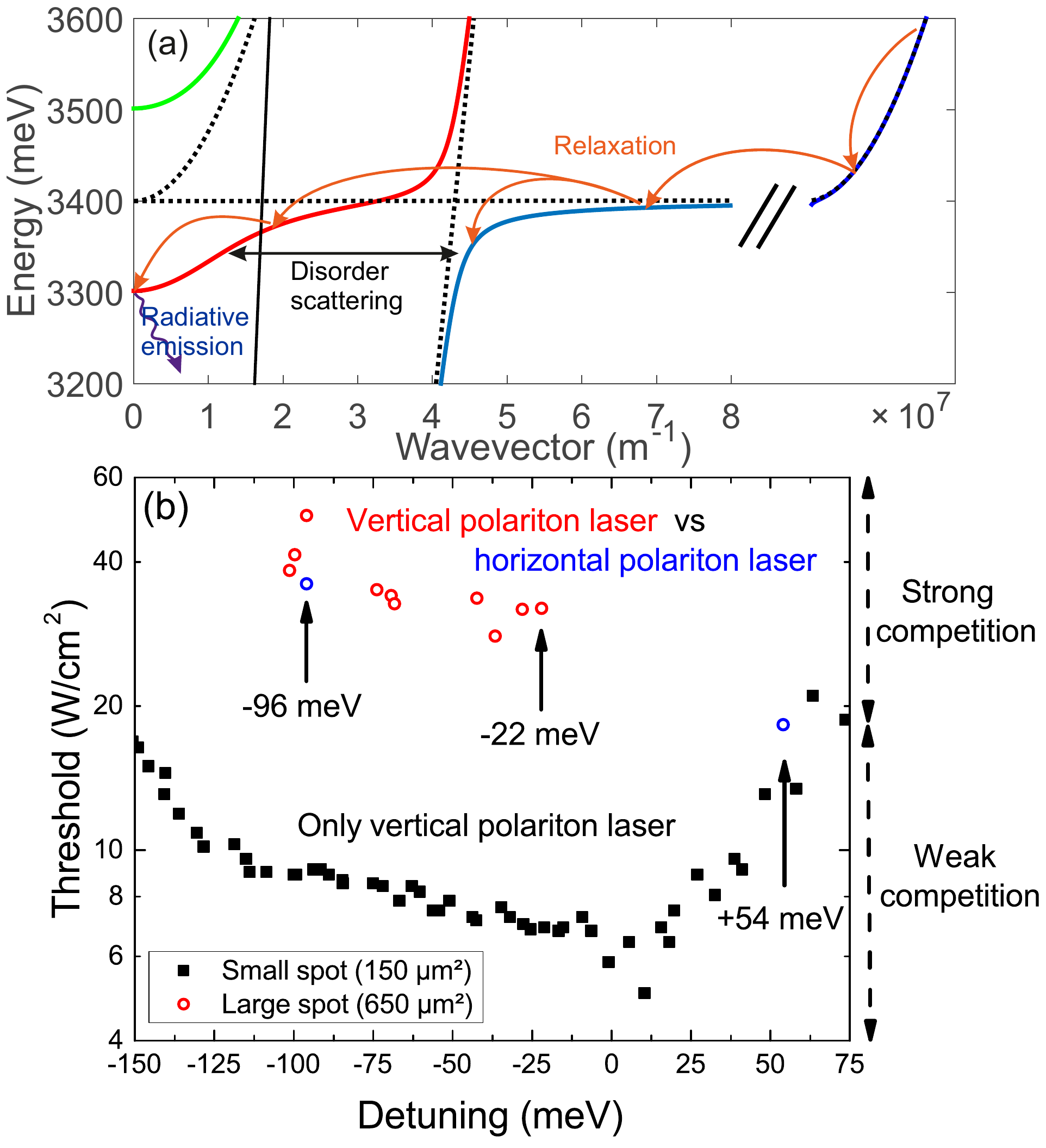}
 \caption{\label{fig1} (a): Polariton dispersion computed using a three coupled oscillators model. Dotted lines: bare cavity photon mode, bare guided photon mode, bare exciton mode. Blue: lower polariton branch (fully guided). Red: Middle polariton branch, which is radiative at low wave vectors. Green: Upper polariton branch. Solid black line: light cone. The arrows show the scattering processes for excitons and polaritons.
(b) Versus exciton-photon detuning, the threshold of the vertical polariton laser measured using a small excitation spot of 150 $\mu$m$^2$ (black square, data taken from \cite{Jamadi2016}) and a large excitation spot of 650 $\mu$m$^2$ (red points). The blue points show the threshold values of the horizontal polariton laser under a large excitation spot. The three arrows indicate the detunings $\delta$ at which the measurements shown in Figs.~2-4 were performed.
}
  \end{center}
 \end{figure}

As shown in many previous works \cite{Kasprzak2008,Boeuf2000,Levrat2010,Lagoudakis2010,
Slootsky2012,Tempel2012,Yamamoto2013,Feng2013},
including Ref.\cite{Jamadi2016} that shows experimental measurements performed on the same sample as in the present work, the polariton lasing threshold in the radiative modes of the cavity depends on the temperature and on the exciton-photon detuning at zero wave vector, defined as  $\delta=E_c-E_x $. The black squares in Fig.~1(b) are extracted from Ref.\cite{Jamadi2016}. They display the measured vertical polariton lasing threshold versus  $\delta$, obtained with an optimum excitation spot of 150~$\mu$m$^2$ at the center of the patterned mesas (i.e. far from the mesa borders, contrary to the present study). For such spot size, we did not find any evidence of horizontal polariton lasing. The detuning dependence shows the usual U-shape, with a range of optimal detuning values minimizing the threshold.

The red points in Fig.~1(b) show the vertical lasing thresholds measured with a larger pumping spot of 650~$\mu$m$^2$. The measured thresholds are three to five times larger than for the smaller spot, due to the enhanced leakage into the guided modes and the competition with the onset of horizontal polariton lasing, as we show in details below. The three arrows in Fig.~1(b) show the detunings at which the measurements presented in the three next figures are performed. For the most negative detuning, polariton lasing in both horizontal and vertical modes is observed with similar thresholds. In the trade-off region (optimal detuning case), only vertical polariton lasing is observed, while at positive detuning, where the vertical lasing threshold is expected to strongly increase, only horizontal polariton lasing is achieved.

\section{Coexistence of horizontal and vertical polariton lasing}

Figure~2 shows the real and reciprocal space PL of the cavity for the most negative detuning $\delta=-96$~meV. Besides this negatively-detuned mode at 3240~meV, a second polariton mode is visible close to 3350~meV (positive detuning), but it is not playing a crucial role. In this case, both vertical and horizontal polariton lasing occur with comparable thresholds. At low pumping (Fig.~2(a)), only the radiative polariton modes are visible. At 35~W/cm$^2$ (Fig.~2(b)) a broad, non-dispersive line appears between both LPBs that comes from guided polaritons essentially  through the mesa edge. At 39~W/cm$^2$ (Fig.~2(c)), multimode lasing takes place from the guided polariton mode band. At 61~W/cm$^2$ (Fig.~2(d)), lasing takes place both in the guided and vertical polariton modes. The corresponding real space image of the emission is shown in Fig.~2(f). The mesa edge is shown by a white dotted line. The pumping laser covers the whole area, being maximum near the vertical laser emission spot. The overall emission clearly comes from two distinct areas in space. Its energy is correlated with the emission spot. The emission below the pump maximum takes place at the radiative polariton energy and corresponds to the vertical polariton laser. The emission coming from the mesa edge is at the energy of the guided modes and corresponds to the horizontal polariton laser. In fact, horizontal polariton lasing certainly takes place everywhere below the excitation laser, but the corresponding emission is extracted out of the sample only at the mesa edge or through structural defects (e.g. cracks). 
Both types of lasers necessarily share the same excitonic reservoir and are therefore competing for it. This conclusion is further supported by Fig.~2(e), which shows the PL spectra as a function of pumping intensity, and by Fig.~2(g), which presents the integrated emission intensity versus pumping for the two types of lasers. The threshold of the horizontal laser is  smaller but it saturates exactly when the vertical laser is reaching its own threshold. 

\begin{figure}[tbp]
 \begin{center}
 \includegraphics[width=1\linewidth]{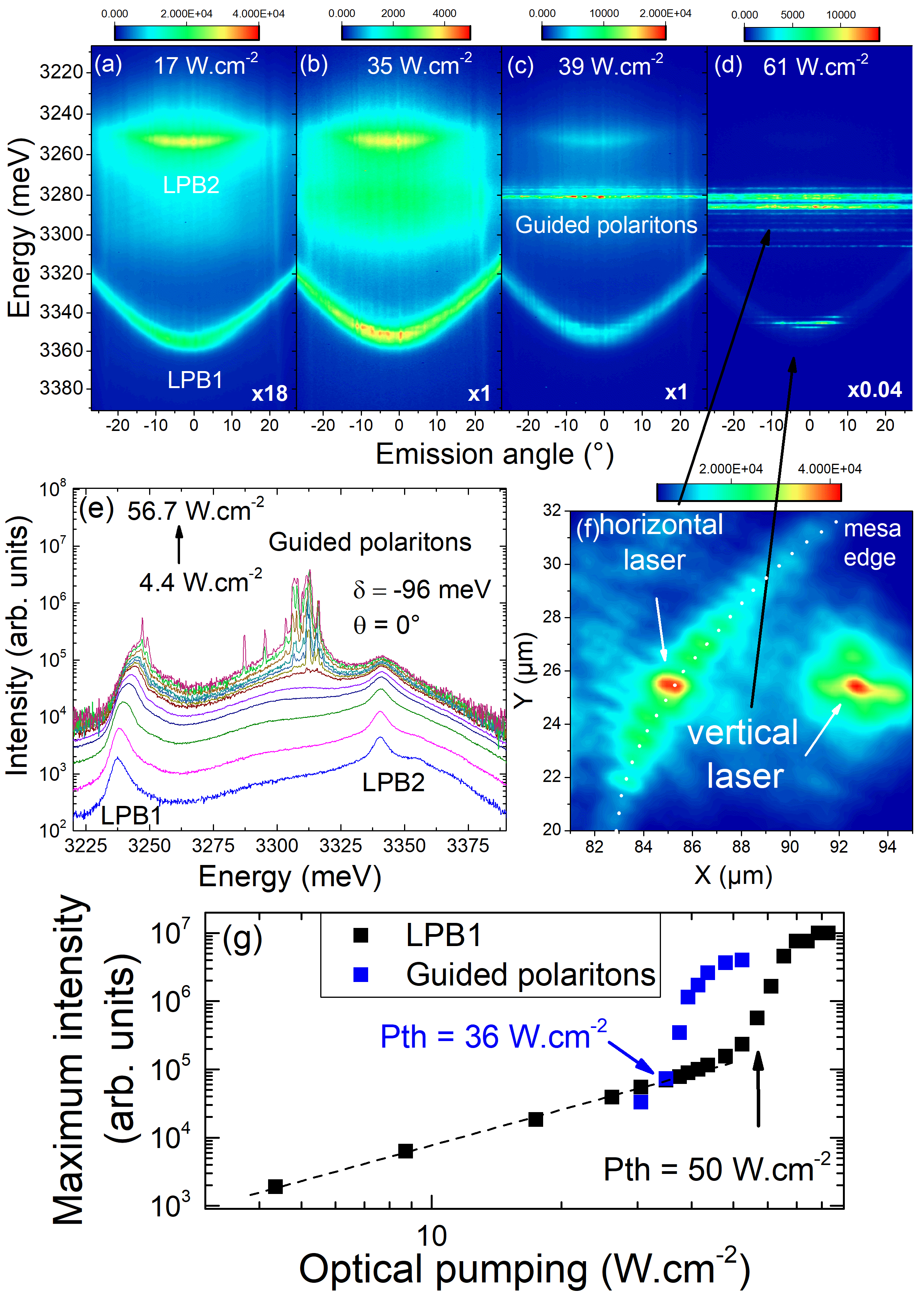}
 \caption{\label{fig2} 
Measurements performed at 5 K and  $\delta= -96$~meV. 
 (a-d): Angularly resolved emission spectrum of the microcavity measured at (a) 17~W/cm$^2$, (b) 35~W/cm$^2$, (c) 39~W/cm$^2$, (d) 61~W/cm$^2$.
 (e) Emission spectra for varied pumping powers: 4, 8.7, 17, 26.2, 30.5,  37.5, 39.2, 41.4, 43.6, 48, 53, 61~W/cm$^2$.
 (f) Real space emission taken at 61~W/cm$^2$. The dotted white line shows the position of the sample edge. The emission from the edge spectrally corresponds to the horizontal polariton laser energy (3312~meV), whereas the energy under the pumping spot corresponds to the one of the vertical polariton laser (3247~meV).
 (g) Emission intensity versus pumping of the guided polariton laser (blue) and of the vertical polariton laser (black).}
  \end{center}
 \end{figure}

One could argue that we cannot prove by this sole experiment that the guided modes are exciton-polariton modes and not just purely photonic states, given that we do not monitor their dispersion. First, one should take into account that we do resolve the dispersion of the vertical-cavity radiative polaritons (Fig.~2(a-d)), which remains almost unperturbed in the whole pumping power range we use. This demonstrates that in this pumping range the strong coupling is maintained. Second, we recently demonstrated horizontal polariton lasing \cite{Jamadi2017} in a ZnO-based waveguide sample, where the dispersion was measured via a grating, and where strong coupling was clearly maintained at least for pumping densities comparable with the ones used in the present set of experiments. The goal of the present work is not to demonstrate horizontal polariton lasing (which we did previously), but to show that this phenomenon can also take place in planar cavities concomitantly with more standard vertical polariton lasing, with which it competes. The fact that guided modes are essentially invisible, except when large defects are present or in samples specially designed to extract such modes, makes this lasing mode difficult to detect and probably explains why it was not reported before. At the detuning used in Fig.~2, the lasing thresholds of both types of lasers are quite similar. One can also note that in Ref.\cite{Jamadi2017} we found a horizontal polariton lasing threshold at 5~K one order of magnitude smaller than in the present work. There are several reasons for this. First, in \cite{Jamadi2017} it was possible to optimize the spatial overlap between the pump and the horizontal cavities, which were defined by vertical cracks, which really allowed to lower the horizontal polariton lasing threshold. Second, the waveguide of \cite{Jamadi2017} was monomode by design, contrary to the current situation, which reduces the extra mode competition that is necessarily present in our 2-$\lambda$ microcavity. Finally, the structural quality of the monolithic active regions in Ref. \cite{Jamadi2017} led to an excitonic nonradiative lifetime three to five times larger than in the current ZnO cavity, which is grown on a nitride distributed Bragg reflector itself grown on top of a silicon substrate. In the following, we will show that by changing $\delta$ it is possible to tune the vertical polariton lasing threshold and favor one or another type of lasing geometry.

Figure~3 shows PL emission taken at an exciton-photon detuning close to the optimum for vertical polariton lasing. The threshold is at about 32~W/cm$^2$. This value is close to the threshold measured for the horizontal polariton laser in Fig.~2. The lasing peak emerges from the upper tail of the lower polariton mode. One should note that some amount of light is certainly scattered by imperfections from the LPB towards the guided modes, which are energetically resonant. It certainly explains why above threshold (Fig.~3(d)), the image in Fourier space contains both the emission coming from the LPB and the dispersiveless features associated with the guided polariton modes.
 
Figure~4 is taken at a sufficiently positive detuning so that the vertical polariton lasing is difficult to reach.  Horizontal polariton lasing occurs with a threshold of 18~W/cm$^2$ for this specific location on the sample. This low value is probably due to the weaker competition with the vertical laser at this detuning. The non-dispersive modes emerge just below the LPB. Figure~4(c,d) shows the spatial and spectral distribution of the emission taken at  threshold. The LPB emission takes place near $X,Y=0$, below the pumping spot, and is centered at 3330~meV. Light is also emitted at the mesa edge ($Y=20~\mu$m), at an energy of 3317~meV. These last features are spectrally  very narrow, typical for lasing, and are weakly visible everywhere in the sample (Fig.~4(d)) due to the existence of scattering mechanisms (e.g. disorder scattering, as depicted in Fig.~1) that allow to partially extract the light from the guided modes into the radiative LPB modes.
  
 \begin{figure}[tbp]
 \begin{center}
 \includegraphics[width=1\linewidth]{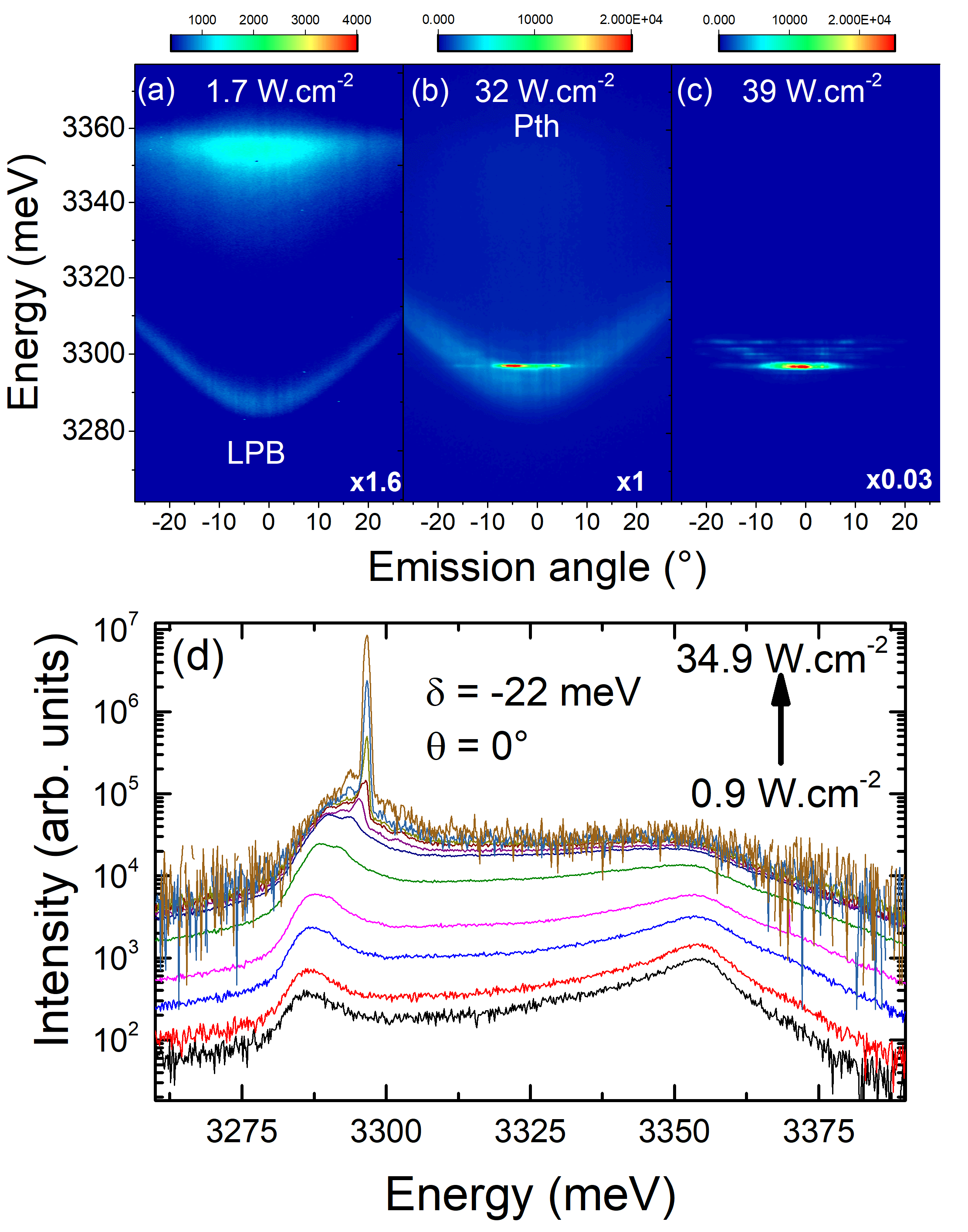}
 \caption{\label{fig3} 
 Measurements performed at 5 K and  $\delta=-22$~meV. 
 (a-c): Angularly resolved emission spectra of the microcavity measured at (a) 1.7~W/cm$^2$, (b) 32~W/cm$^2$, (c) 39~W/cm$^2$.
 (d) Emission spectra for varied pumping powers, 1, 1.7, 4.4, 8.7, 17.4, 26.2, 30, 31.4, 32.3, 33, 35~W/cm$^2$. 
}
  \end{center}
 \end{figure}

\begin{figure}[tbp]
 \begin{center}
 \includegraphics[width=1\linewidth]{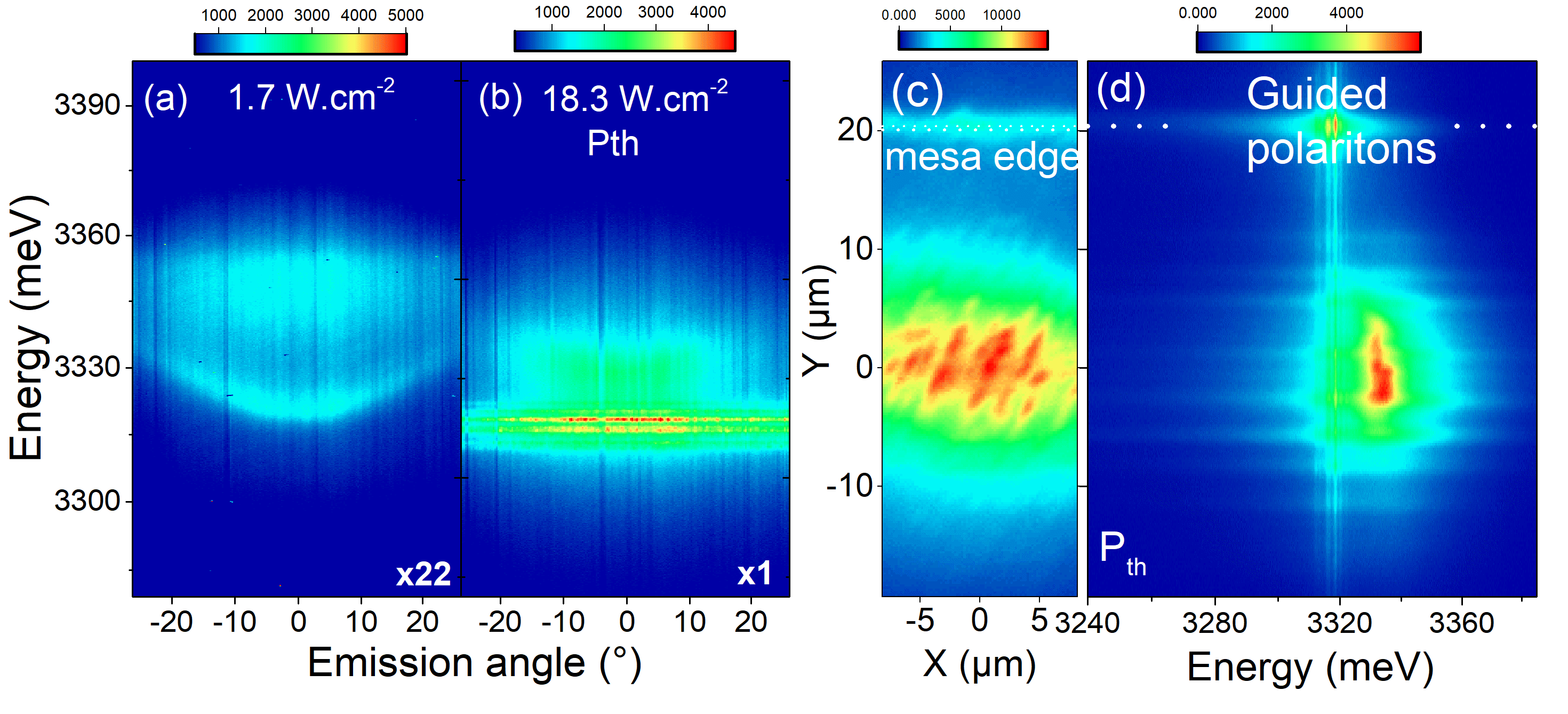}
 \caption{\label{fig4} 
Measurements performed at 5~K and  $\delta=+54$~meV
 (a-b): Angularly resolved emission spectrum of the microcavity measured at (a) 1.7~W/cm$^2$, (b) 18.3~W/cm$^2$.
 (c) Spatially resolved emission (X,Y) at 18.3~W/cm$^2$. The white dotted line at $Y=20~\mu$m shows the sample edge.
 (d) Spatially (Y-Axis) and energetically resolved emission taken at 18.3~W/cm$^2$. The edge emission is energetically sharp and distinct from the LPB energy located under the pumping spot centered at $Y=0$.
}
  \end{center}
 \end{figure}


\section{Discussion}
  
It is interesting to note that this sample has been already thoroughly studied as a vertical polariton laser \cite{Jamadi2016}. In that previous study we did not detect the occurrence of horizontal polariton lasing. This is really because the guided modes are weakly visible in such samples, unless one measures either close to a crack or to a mesa border. Furthermore, the cracks in the active region are often covered by the top Bragg mirror, falling into its bandgap, which prevents the emission. Therefore, the light can only get out near the vertical direction in the radiative photonic modes, namely the LPBs. In fact, horizontal lasing, which is very strongly present, is clearly visible only near the mesa edge. 
The measurements are usually performed far from the edges because of the presence of defects and of border effects modifying locally the designed microcavity structure (e.g. due to enhanced growth rates compared to the mesa centers). Moreover, it is in general hard to find something which is not expected.

Several consequences can be deduced from the current findings. First, theoretical calculations of polariton lasing threshold in vertical microcavities always neglected the possibility of lasing taking place in the guided modes. Our findings can really explain quantitative discrepancies between theory and experiment on the determination of these thresholds. The other consequence, which is much more crucial, is that guided modes, and even lasing in these modes, could have taken place without being noticed in a very large number of previous reports of vertical polariton lasing. It is in that sense that our present report is essential. This is of course especially true for cavities showing moderate $Q$, probably below $10^4$, such as the GaAs cavities studied in the 90's and the 2000's \cite{Lagoudakis2004}, and for the CdTe-based cavities, where the first BEC effect was reported \cite{Kasprzak2006}. It is also the case for all structures where room temperature polariton lasing was observed, namely GaN, ZnO, organics, and perovskite-based cavities. Only the latest GaAs based structures with $Q$-factors of $10^5$ or larger \cite{Sun2017,Estrecho2018} seem to be relatively well protected from competition with guided modes. However, even in this last case etched structures could show confined guided modes that could be living longer in the pump area than the radiative modes. One could also note that the "polariton condensate instability" is often observed when large pumping spots are used \cite{Baboux18,Kena-cohen}, which is also the regime favorable for horizontal lasing, an aspect completely neglected by all models so far.

To conclude, we demonstrate that horizontal polariton lasing in guided modes and vertical polariton lasing in radiative modes can simultaneously take place in a planar microcavity. This is a crucial finding that could have affected for quite a long time many polariton lasing and polariton condensation reports. It might be impacting both polariton device operation and, from a more fundamental point of view, polariton condensate features.

\begin{acknowledgments}
We acknowledge the support of the ANR projects: "Plug and Bose" (ANR-16-CE24-0021), "Quantum Fluids of Light" (ANR-16-CE30-0021) and the "Investissement d'avenir" program GANEX (ANR-11-LABX-004), IMOBS3 (ANR-10-LABX-16-01), ISITE "Cap2025" (16-IDEX-0001). C2N is a member of RENATECH (CNRS), the national network of large micro-nanofabrication facilities. D.D.S. acknowledges the support of IUF (Institut Universitaire de France).
\end{acknowledgments}

\bibliography{reference}

\end{document}